\documentclass[12pt,a4paper]{article}
\usepackage[left=2cm,top=2cm,right=2cm,bottom=2.5cm,foot=2cm, nohead]{geometry}

\usepackage{graphicx}
\usepackage{epsfig}
\usepackage{amsmath,amsfonts,amssymb,amsthm}
\usepackage{color}
\usepackage{hyperref}
\usepackage[sort]{natbib}
\usepackage{hyperref}
\hypersetup{colorlinks=true, citecolor=green}
\usepackage{setspace,enumitem}
\usepackage{color}
\usepackage{fancyvrb}
\usepackage{caption}
\usepackage{array}
\usepackage[table]{xcolor}
\usepackage[thinlines]{easytable}
\usepackage{arydshln}

\newcommand{\BEAS}{\begin{eqnarray*}}
\newcommand{\EEAS}{\end{eqnarray*}}
\newcommand{\BEA}{\begin{eqnarray}}
\newcommand{\EEA}{\end{eqnarray}}
\newcommand{\BIT}{\begin{itemize}}
\newcommand{\EIT}{\end{itemize}}
\newcommand{\BNUM}{\begin{enumerate}}
\newcommand{\ENUM}{\end{enumerate}}

\newcommand{\NN}{\nonumber}

\newcommand{\sign}{{\rm sign}}

\newtheorem{theorem}{Theorem}

\newtheorem{assumption}{Assumption}

\newtheorem{example}{Example}

\title{A general class of quasi-independence tests for left-truncated right-censored data}
\date{\today}
\author{Young-Geun Choi, Wei-Yann Tsai and Myunghee Cho Paik\thanks{Young-Geun Choi is Postdoctoral Fellow, Public Health Sciences Division, Fred Hutchinson Cancer Research Center, Seattle, WA, USA. Wei-Yann Tsai is Professor, Department of Biostatistics, Columbia University, New York, NY, USA. Myunghee Cho Paik is Professor, Department of Statistics, Seoul National University, Seoul, Korea. All correspondences are to Myunghee Cho Paik(\texttt{myungheechopaik@snu.ac.kr}).}}
\date{\today}

\begin{document}

\maketitle

\begin{abstract}
In survival studies, classical inferences for left-truncated data require quasi-independence, a property that   the joint density of truncation time and failure time is factorizable into their marginal densities in the observable region.
The quasi-independence hypothesis is testable; many authors have developed tests for left-truncated data with or without right-censoring. 
In this paper, we propose a class of test statistics for testing the quasi-independence which unifies the existing methods and generates new useful statistics such as conditional Spearman's rank correlation coefficient. 
Asymptotic normality of the proposed class of statistics is given.
We show that a new set of tests can be powerful under certain alternatives by theoretical and empirical power comparison.
\\

\noindent {\bf Keywords:} Conditional Kendall's tau; Conditional product-moment correlation; Left-truncation; Quasi-independence testing; Right-censoring; Survival data.
\end{abstract}

\baselineskip 22pt

\section{Introduction}\label{sec:info}

Left truncation is a common type of incompleteness in survival analysis along with right-censoring.   A typical prevalent cohort study produces left-truncated data.  A famous example is the study of the residents of the Channing House retirement community in Palo Alto, California.  The subjects who died before becoming eligible for the retirement community cannot be recruited and thus are truncated. Left truncation also induces length-biased sampling \citep{Asgharian2002,Tsai2009}. Analysis of left-truncated data with or without right censoring  has been studied \citep{Lynden-Bell1971,Hyde1977,Wang1986}.  To accommodate left truncation, a key is to redefine the risk set for truncated data to apply the Kaplan-Meier estimator and Cox proportional hazard models \citep{Andersen1993}.  However, these methods assume an independence between failure time and truncation time.  \cite{Tsai1990} pointed out that what we need is a weaker assumption than independence, namely, quasi-independence,  which is independence between failure and truncation times only in the observed region.

Unlike independence between failure time and censoring time, quasi-independence is testable. \cite{Tsai1990} proposed a statistic based on the Kandall's tau conditioning on a set of observations that can be interpreted as  ``comparable'' set of data points. \cite{Jones1992} presented a score-type test from the Cox proportional hazard model. \cite{Chen1996} suggested a test statistic for quasi-independence based on the conditional product-moment correlation among comparable data points  if failure and truncation times follow truncated bivariate normal distribution. \cite{Martin2005} extended the conditional Kendall's tau test to more complex truncation schemes. We note that both \cite{Tsai1990} and \cite{Chen1996}'s works adapted usual association measures for complete bivariate data. It may become of interest whether other measures can be adjusted for left truncation and right censoring. However, the following questions remain unclear in general: (1) which association measure for complete bivariate data can be tailored to survival data with left truncation?; (2) if so, in what mechanism the given measure is adjusted?


To address those question, we propose a general class of test statistics  that unifies the existing tests and induces some new statistics. The class reveals that any association measure with a form $U$-statistic comprising skew-symmetric transformations can induce a version for left truncated data, restricting the measure onto the mean among ``comparable pairs''. 
The class is broad enough to embed not only existing tests by  \cite{Tsai1990}, \cite{Jones1992} and \cite{Chen1996}, but also Spearman's rank correlation, another well-known association measure.
We derive asymptotic properties of the proposed class of statistics. We further specify sufficient conditions to guarantee that the proposed class has a valid asymptotic null distribution under right censoring, which is another contribution.
We compare various statistics by simulation studies and theoretical efficacies under a sequence of contiguous alternatives.

The remainder of this paper is organized as follows. In Section \ref{sec:existing}, we introduce notations and  review existing quasi-independence tests. In Section \ref{sec:proposal} we propose the general class of test statistics and derive asymptotic properties. We also introduce  new useful  statistics derived from the proposed class. In Section \ref{sec:AREs}, we compare asymptotic relative efficiencies of some cases of our class to Kendall's tau under a special case of contiguous excess and relative risk models.  In Section \ref{sec:simul}, we  compare the empirical power of the different tests within the class under various alternatives.   Section \ref{sec:data} illustrates application of the proposed tests to the Channing House data.   Finally, concluding remarks follow in Section \ref{sec:conclusion}.

\section{Existing methods}\label{sec:existing}

In this Section, we review existing methods for quasi-independence testing: conditional Kendall's tau \citep{Tsai1990}, conditional product-moment correlation coefficient \citep{Chen1996}, and a weighted score test for the Cox proportional hazard model \citep{Jones1992}.

Let  $(L, X, C)$ be a tuple of truncation,  failure, and censoring  times, and  $T=\min(X,C)$ and $\delta=I(X \leq C)$ where $I(A)$ is the indicator function for an event $A$. 
The quasi-independence hypothesis is described by
\begin{equation}\NN
H_0 : \quad {\rm pr}(L \le l, X \le x \,|\, L\le X) = {\rm pr}(L \leq l) {\rm pr}(X \leq x) / \alpha, \quad  l < x
\end{equation}
for some constant $\alpha$.
We suppose that left-truncated data $\{ (L_i, X_i) \}_{i=1}^n$ are independently and identically distributed given $L < X$.  In the presence of right censoring,   left-truncated right-censored data $\{ (L_i, X_i, C_i) \}_{i=1}^n$ are  assumed to be randomly sampled given $L < X$ where we observe $\{ (L_i, T_i, \delta_i) \}_{i=1}^n$.
In addition, we assume throughout the paper that
\begin{assumption}
\label{A1} $X$ and $C$ are independent conditionally on $L$, and
\end{assumption}
\begin{assumption}
\label{A2} $L < C$ with probability 1.
\end{assumption}

\subsection{Conditional Kendall's tau}\label{subsec:ctau}

Kendall's tau is a popular nonparametric measure of association between two random variables.
If $(X_1, Y_1)$ and $(X_2, Y_2)$ are independent copies of a bivariate random vector $(X, Y)$, Kendall's tau is defined by
\[
\rho^{\rm Tau} =  E\big\{ \sign(X_1 - X_2) \sign(Y_1 - Y_2) \big\},
\]
where $\sign(t)$ is the sign function, $\sign(t) = I(t > 0) - I(t < 0)$.
It is known that if $X$ and $Y$ are independent, then $\rho^{\rm Tau} = 0$. A consistent estimator of $\rho^{\rm Tau} $ can be obtained by a $U$-statistic
\[
\widehat{\rho}^{\rm Tau} = \frac{1}{\binom{n}{2}} \sum_{i=1}^{n-1}\sum_{j=i+1}^{n} \sign(X_j - X_i)\sign(Y_j - Y_i).
\]
One can construct a test statistic based on $\widehat{\rho}^{\rm Tau}$ to test $\rho^{\rm Tau}  = 0$, and its limiting distribution can be obtained applying  the central limit theorem for $U$-statistic \citep{Randles1991}.

\citet{Tsai1990} defined conditional Kendall's tau to adjust the Kendall's tau for left-truncated data. Let $(L_1,X_1)$ and $(L_2,X_2)$ be independent copies from the distribution of $(L,X)$ given $L<X$. The conditional Kendall's tau is defined by
\begin{eqnarray}
\rho^{\rm CTau} &=& E\big\{ \sign(L_1 - L_2) \sign(X_1 - X_2) | \Omega_{12} \big\} \NN \\
	 &=& E\big\{ \sign(L_1 - L_2) \sign(X_1 - X_2) I(\Omega_{12}) \big\} / {\rm pr}(\Omega_{12}), \label{eqn:ctau}
\end{eqnarray}
where $\Omega_{12}$ is an event defined by $\Omega_{12} = \{ \max (L_1, L_2) < \min(X_1, X_2) \}$. We call $\Omega_{12} $ comparable region.
In this region, there is overlap between the two durations. 
The conditional Kendall's tau   measures concordance only in comparable region, $\Omega_{12}$. It is easy to show that $\rho^{\rm CTau} = 0$  between $L$ and $X$ under the quasi-independence.  A consistent estimator of $\rho^{\rm CTau}$ (\ref{eqn:ctau}) is 
\begin{eqnarray}\NN
\widehat{\rho}^{\rm CTau} &=&  \left\{ \frac{1}{\binom{n}{2}} \sum_{i=1}^{n-1}\sum_{j=i+1}^{n} \sign(L_j - L_i)\sign(X_j - X_i)I(\Omega_{ij}) \right\} \Big/  \left\{ \frac{W_n}{\binom{n}{2}} \right\} \NN \\
	&=& \frac{1}{W_n} \sum_{i=1}^{n-1}\sum_{j=i+1}^{n} \sign(L_j - L_i)\sign(X_j - X_i)I(\Omega_{ij}) \NN
\end{eqnarray}
where $W_n = \sum_{i=1}^{n-1}\sum_{j=i+1}^{n}I(\Omega_{ij})$.
As in $\widehat{\rho}^{\rm Tau}$, the conditional version $\widehat{\rho}^{\rm CTau}$ is approximated by a normal distribution from the same central limit theorem. The asymptotic variance of $\widehat{\rho}^{\rm CTau}$ can be obtained as the variance of $U$-statistic using $O(n^3)$ of FLoating-point OPerations per Second (flops), where the formula is introduced in the next Section. \cite{Martin2005} introduced an analytic technique in which the computation reduces to $O(n^2)$ flops. On the other hand, \cite{Tsai1990} derived another formula for the variance which only involves the numbers of the risk sets of given data and requires at most $O(n^2)$ flops.

Conditional Kendall's tau  was adapted to right-censoring by \cite{Tsai1990}, and \cite{Martin2005} in multivariate case. Having $(L_1,T_1,\delta_1)$ and $(L_2,T_2,\delta_2)$ as independent copies conditioning on $L<X$, the quantity of interest is
\begin{eqnarray}
{\kappa}^{\rm CTau} &=& E\big\{ \sign(L_1 - L_2) \sign(T_1 - T_2) | \Lambda_{12} \big\} \NN \\
	 &=& E\big\{ \sign(L_1 - L_2) \sign(T_1 - T_2) I(\Lambda_{12}) \big\} / {\rm pr}(\Lambda_{12}), \NN
\end{eqnarray}
where 
 $\Lambda_{12}$ is a comparable region with left truncation and right censoring defined by $\Lambda_{12} = \{ \max (L_1, L_2) < \min(T_1, T_2) \}$ $\cap$  $[ (\delta_1 \delta_2 = 1)$ $\cup$ $\{\delta_1 \, \sign (T_2 - T_1) = 1\}$ $\cup$  $\{\delta_2 \, \sign (T_1 - T_2) = 1 \}] $.
Due to right censoring, comparable region $\Lambda_{12}$ is modified from $\Omega_{12}$ so that minimum of $T_1$ and $T_2$ should be a failed one.  
As in $\Omega_{12}$, there is overlap between the two durations in $\Lambda_{12}$.  
If $L$ and $X$ are quasi-independent, \cite{Martin2005} proved that ${\kappa}^{\rm CTau} = 0$ under Assumptions \ref{A1} and \ref{A2}. A consistent estimator of $\kappa^{\rm CTau}$ is 
\begin{equation}\label{eqn:hatctau2}
\widehat{\kappa}^{\rm CTau} =
	\frac{1}{V_n} \sum_{i=1}^{n-1}\sum_{j=i+1}^{n} \sign(L_j - L_i)\sign(T_j - T_i)I(\Lambda_{ij}),
\end{equation}
where $V_n = \sum_{i=1}^{n-1}\sum_{j=i+1}^{n}I(\Lambda_{ij})$.

\subsection{Conditional product-moment correlation}\label{subsec:cprod}

\cite{Chen1996} proposed a test based on conditional product-moment correlation coefficient for left-truncated data. Similarly to \cite{Tsai1990}, \cite{Chen1996} first defined the population version of the correlation coefficient conditioning on $\Omega_{12}$,
\begin{equation}\NN
\rho^{\rm CProd} \,=\, E\big\{ (L_1 - L_2)(X_1 - X_2) | \Omega_{12} \big\},
\end{equation}
and proved $\rho^{\rm CProd}=0$ under the quasi-independence. They also suggested a consistent estimator
\begin{equation}\NN
\widehat{\rho}^{\rm CProd} =
	\frac{1}{W_n} \sum_{i=1}^{n-1}\sum_{j=i+1}^{n} (L_j - L_i)(X_j - X_i)I(\Omega_{ij}).
\end{equation}

It is noteworthy that \cite{Chen1996} did not extend the test to the right-censored data. As in the case of the conditional Kendall's tau, one might have considered an analogue of $\rho^{\rm CProd}$ for right-censoring, for example, say $\kappa^{\rm CProd}  = E\big\{ (L_1 - L_2)(T_1 - T_2) |\Lambda_{12} \big\}$.  However, in general, $\kappa^{\rm CProd} \neq 0$ under the quasi-independence, and a similar extension with conditional Kendall's tau fails. 
In Section \ref{sec:proposal}, we will see that $\kappa^{\rm CProd}=0$ holds with additional assumptions.


\subsection{Weighted score tests from the Cox proportional hazard model}\label{subsec:coxph}

\cite{Jones1992} adopted a class of weighted score tests from the Cox proportional hazard model proposed in \cite{Jones1989} to the quasi-independence testing.
Consider a conditional hazard function with a time-varying covariate given by
\begin{equation}\NN
\lambda(t \,|\, Z_i ) = \lambda_0 (t) \, \exp \left\{ Z_i(t)\beta \right\}, \quad i = 1, \ldots, n ,
\end{equation}
where $\lambda(t\,|\,Z)$ is the hazard function for $X$ given a predictable covariate function $Z(t)$ and the truncation event $L<X$. Testing $\beta = 0$ yields a score-type test statistic and \cite{Jones1989} proposed its weighted version,
\begin{equation}\NN
 T^{\rm Cox}(q, Z) =
  \sum_{i=1}^{n} \delta_i \, q(T_i) \left\{ Z_i(T_i) - \frac{ \sum_{j=1}^{n} Y_j(T_i) Z_j(T_i) }
	 { Y(T_i) } \right\},
\end{equation}
where $q(t)$ is a predictable weight function, $Y_j(t) = I(L_j < t \leq T_j)$ is the at-risk indicator for the $j$-th observation at time $t$, and $Y(t) = \sum_{j=1}^{n} Y_j(t)$ is the number of subjects at risk for time $t$. The asymptotic distribution of $T^{\rm Cox}(q, Z)$ under $\beta=0$ is derived using the counting process theory.
\cite{Jones1992} proposed to take a function of truncation time as covariate.   Then the quasi-independence  is equivalent to $\beta=0$, which can be tested via the score test,   $T^{\rm Cox}$.      The covariates considered in the paper were $Z_i(t) = L_i$ or $Z_i(t) = R_i^*(t)$, where $R_i^*(t) = R_i(t) / Y(t)$ and $R_i(t)$ is the rank of $L_i$ in the risk set defined at $t$, explicitly, $R_i(t) =  1 + \sum_{j=1}^{n} Y_j(t) I(L_i < L_j) $.   We  denote  the weighted  score statistics with covariate $L_i$ or $R_i^*(t)$ by  $T^{\rm Cox}(q, L)$ and  $T^{\rm Cox}(q, R^*)$, respectively.

\section{A general class for quasi-independence tests}\label{sec:proposal}

A common characteristic from the conditional Kendall's tau and the conditional product-moment correlation
is that they have a form of $U$-statistics among the pairs in a comparable region defined in the presence of left truncation and right censoring. We begin this Section by observing that the tests from the weighted score tests from the Cox proportional hazard model also can be expressed as a $U$-statistic. Considering $q(t)=Y(t)$, the size of risk set at time $t$ as the weight function, we have
\begin{equation}\label{eqn:coxU1}
T^{\rm Cox}\{Y, a(L)\} = - \sum_{i=1}^{n-1} \sum_{j=i+1}^{n} \{ a(L_i) - a(L_j) \}\sign(T_i - T_j)I(\Lambda_{ij}) 
\end{equation}
for any real-valued function $a(\cdot)$ and
\begin{equation}\label{eqn:coxU2}
T^{\rm Cox}(Y, R^*) = - \frac{1}{2} \sum_{i=1}^{n-1} \sum_{j=i+1}^{n} \sign(L_i - L_j)\sign(T_i - T_j)I(\Lambda_{ij}) + \frac{1}{2}\sum_{i=1}^{n} \delta_i.
\end{equation}
The derivation of \eqref{eqn:coxU1} and \eqref{eqn:coxU2} is given in the Appendix.
\cite{Jones1992} discuss the relationship in \eqref{eqn:coxU2}, i.e., $T^{\rm Cox}(Y, R^*)$ is equivalent to the conditional Kendall's tau. From the two equations, we see that $T^{\rm Cox}(Y, L)$ and $T^{\rm Cox}(Y, R^*)$ share the common characteristic of $\widehat{\rho}^{\rm CTau}$, $\widehat{\kappa}^{\rm CTau}$, and $\widehat{\rho}^{\rm CProd}$, namely,  $U$-statistic indexed  among comparable pairs adjusting left truncation and right censoring.

Another important common characteristic among the statistics, $\widehat{\rho}^{\rm CTau}$,  $\widehat{\kappa}^{\rm CTau}$,   $\widehat{\rho}^{\rm CProd}$, $T^{\rm Cox}(Y, L)$, and $T^{\rm Cox}(Y, R^*)$,  is  skew-symmetry of the functions of $L_i$ and $L_j$, satisfying $g(L_i,L_j)=-g(L_j,L_i)$.  To see why the skew-symmetry of $g$ is crucial, let us consider the case of left-truncation without right-censoring. For all those $U$-statistics, we need their estimands to be zero under quasi-independence to have valid asymptotic null distribution.  In taking the expectation, the integral region $\Omega_{12}$ can be divided by four subevents, $A_1 = ( L_1 < L_2 < X_1 < X_2  )$, 
$A_2 = ( L_1 < L_2 < X_2 < X_1  )$,
$A_3 = ( L_2 < L_1 < X_1 < X_2  )$,
and $A_4 = ( L_2 < L_1 < X_2 < X_1  )$, flipping the order of $L_1$ and $L_2$, and $X_1$ and $X_2$, under the restriction of comparable region. Then the expectation can be expressed as a sum of four subintegrals associated with the four regions. The skew-symmetry of the mapping function for truncation times renders cancellation of subintegrals from $A_1$ and $A_3$, and similarly from $A_2$ and $A_4$. 

Motivated by the common structure of the  existing statistics, we propose a class of $U$-type test statistics that is indexed by skew-symmetric functions. We first present the proposed method  for the left-truncated data in Section \ref{subsec:proposed} and then for left-truncated and right-censored data in Section \ref{subsec:proposed2}. The cancellation of subintegrals will play a key role when proving the asymptotic null distribution of the classes. In Section \ref{subsec:examples}, we discuss several new tests that are special cases of the proposed class of statistics, including a version of Spearman's rank correlation coefficient.

\subsection{Tests for left-truncated data without right-censoring}\label{subsec:proposed}

We assume $\{(L_i, X_i)\}_{i=1}^n$ are observed in region  $\{(l,x): l<x \}$.  Let  $g$ and $h$ be bivariate skew-symmetric functions, that is, $g(s,t) = -g(t,s)$ and $h(s,t) = -h(t,s)$ for all real $s, t$.  Skew-symmetry of $g(\cdot,\cdot)$ is needed to cancel out subintegrals from subregions as described earlier.  We impose skew-symmetry of $h(\cdot,\cdot)$ to satisfy symmetry of $U$-statistics.
 Recall the definition $\Omega_{ij} = ( \max (L_i, L_j) < \min(X_i, X_j))$.   We propose
\begin{equation}\label{eqn:hatproposed}
\widehat{\rho}(g, h) = \frac{1}{ W_n } \sum_{i=1}^{n-1}\sum_{j=i+1}^{n}  g(L_i, L_j) h(X_i, X_j) I(\Omega_{ij}),
\end{equation}
where $W_n = \sum_{i=1}^{n-1}\sum_{j=i+1}^{n}I(\Omega_{ij})$, which estimates
\begin{eqnarray}
\rho (g, h) &=& E\big\{ g(L_1, L_2) h(X_1, X_2) | \Omega_{12} \big\} \NN \\
	 &=& E\big\{ g(L_1, L_2) h(X_1, X_2) I(\Omega_{12}) \big\} / {\rm pr}(\Omega_{12}). \label{eqn:proposed}
\end{eqnarray}
The special cases of $\rho (g, h)$ include  $\rho^{\rm CTau}$, by choosing $g(s,t) = h(s,t) = \sign (s-t)$ and $\rho^{\rm CProd}$,   $g(s,t) = h(s,t) =s-t$, respectively. Other choices of $g$ and $h$ generate new statistics, which we discuss in Section \ref{subsec:examples}.


\begin{theorem}\label{thm1}
Let $g(\cdot, \cdot)$ and $h(\cdot, \cdot)$ be skew-symmetric bivariate functions and let $\widehat{\rho}(g, h)$ and $\rho(g, h)$ be defined as in \eqref{eqn:hatproposed} and \eqref{eqn:proposed}. As  $n \rightarrow \infty$,  
$
 n \widehat{\rho}(g, h)^2 /  \{ 4 \phi(g,h) / {\rm pr}(\Omega_{12})^2 \} 
$ is asymptotically distributed as a chi-squared distribution with 1 degree of freedom  under the quasi-independence between $L$ and $X$, where
 $\phi(g,h)=E \{ g(L_1, L_2) h(X_1, X_2)$ $ I(\Omega_{12}) g(L_1, L_3) h(X_1, X_3) I(\Omega_{13}) \}$, and $(L_i, X_i)$, $i=1, 2, 3$ are  independent copies of the distribution of $(L,X)$ conditioning on $L<X$. 
\end{theorem}
A proof  is given in the Appendix.  A key step in the proof is showing that
$\rho (g, h) = 0$ under the quasi-independence between $L$ and $X$ using skew-symmetry of function $g$, following by
  the central limit theorem for one-sample $U$-statistic \citep{Randles1991}.   We can replace 
 ${\rm pr}(\Omega_{12})$ and $\phi(g,h)$ with 
$\widehat{\rm pr}(\Omega_{12}) = \sum_{i<j}I(\Omega_{ij})/\binom{n}{2}$ and
$\widehat{\phi}(g,h) = \sum_{i<j<k} a_{ij} b_{ik} / \binom{n}{3} $, where $a_{ij} = b_{ij} =  g(L_i, L_j) h(X_i, X_j) I(\Omega_{ij})$, and the Theorem  still holds after applying Slutsky's theorem.
 \cite{Martin2005}  reduced the burden of computation of $\widehat{\phi}(g,h)$  with $O(n^3)$ flops to  $O(n^2)$ flops  with
\begin{equation}\label{eqn:approx}
\sum_{i=1}^{n}\sum_{j \neq i}\sum_{k \neq i} a_{ij} b_{ik} =
	\sum_{i=1}^{n} ( a_{i\cdot}b_{i\cdot} - c_{i\cdot}),
\end{equation}
where $a_{i\cdot} =  \sum_{j \neq i} a_{ij}$, $b_{i\cdot} =  \sum_{j \neq i} b_{ij}$, and $c_{i\cdot} = \sum_{j \neq i} a_{ij} b_{ij}$. 

\subsection{Tests for left-truncated right-censored data}\label{subsec:proposed2}


\noindent Given $\{(L_i, T_i, \delta_i)\}_{i=1}^n$, we propose 
\begin{equation}\NN
\widehat{\kappa}(g, h) = \frac{1}{ V_n } \sum_{i=1}^{n-1}\sum_{j=i+1}^{n}  g(L_i, L_j) h(T_i, T_j) I(\Lambda_{ij}),
\end{equation}
where $V_n = \sum_{i=1}^{n-1}\sum_{j=i+1}^{n}I(\Omega_{ij})$ and $\Lambda_{ij} = \{ \max (L_i, L_j) < \min(T_i, T_j) \}$ $\cap$  $[ (\delta_i \delta_j = 1)$ $\cup$ $\{\delta_i \, \sign (T_j - T_i) = 1\}$ $\cup$  $\{\delta_j \, \sign (T_i - T_j) = 1 \}] $. Technically speaking, $\widehat{\rho} (g, h)$ is a special case of $\widehat{\kappa} (g, h)$ with assigning $\delta_i = 1$ for all $i$.
Similarly to the case of $\widehat{\rho}(g,h)$, $\widehat{\kappa}(g, h)$ consistently estimates
\begin{equation}\NN
\kappa (g, h) = E\big[ g(L_1, L_2) h(T_1, T_2) I(\Lambda_{12}) \big] / {\rm pr}(\Lambda_{12}).
\end{equation}

In addition to Assumptions \ref{A1} and \ref{A2}, we suppose the following:
\begin{assumption}
\label{A3}
Either (3A) $h(s,t) = \sign(s-t)$ or (3B) $L$ and $C$ are quasi-independent.
\end{assumption}
Then we have the following theorem:

%
\begin{theorem}\label{thm2}
Let $g(\cdot, \cdot)$ and $h(\cdot, \cdot)$ be skew-symmetric bivariate functions and Assumption \ref{A3} holds.  As $n \rightarrow \infty$, 
$ n \widehat{\kappa}(g, h)^2 / \{ 4 \varphi(g,h) / {\rm pr}(\Lambda_{12})^2 \} $ is asymptotically distributed as a chi-squared distribution with 1 degree of freedom under the quasi-independence between $L$ and $X$, where 
$\varphi(g,h) = E\left\{ g(L_1, L_2) h(T_1, T_2) I(\Lambda_{12}) g(L_1, L_3) h(T_1, T_3) I(\Lambda_{13}) \right\}$,  and  $(L_i, T_i, \delta_i)$, $i=1, 2, 3$ are  independent copies of the distribution from $(L,T,\delta)$ conditioning on $L<X$. 
\end{theorem}
A sketch of a proof of is given in the Appendix.   As in Theorem \ref{thm1}, a key idea is to show  $\kappa (g, h) = 0$ under the quasi-independence between $L$ and $X$ 
given Assumption \ref{A3} in addition to \ref{A1} and \ref{A2}.  We can show  that $\kappa (g, h)$ becomes zero under quasi-independence with Assumption 3A using the skew-symmetry of $g(\cdot,\cdot)$.     If $h(\cdot,\cdot)$ is a skew-symmetric function other than the sign function  as in the conditional product-moment correlation, we need extra assumption, Assumption 3B. 
By Slutsky's theorem, we can replace $\varphi(g,h)$ and ${\rm pr}(\Lambda_{12})$ with  respective consistent estimators, $\widehat{\rm pr}(\Lambda_{12}) = \sum_{i<j}I(\Lambda_{ij})/\binom{n}{2}$ and
$\widehat{\varphi}(g,h) = \sum_{i<j<k} a_{ij} b_{ik} / \binom{n}{3} $, where $a_{ij} = b_{ij} =  g(L_i, L_j) h(T_i, T_j) I(\Lambda_{ij})$.


\subsection{New test statistics and a unifying framework}\label{subsec:examples}

We mentioned that the test statistics $\widehat{\rho}(g,h)$ and $\widehat{\kappa}(g,h)$ embed the existing tests described in Section \ref{sec:existing}. 
Other choices of $g$ and $h$ generate several new useful tests. We highlight the following examples:

\begin{example}\label{ex1}
Let $g(L_i, L_j) = L_i - L_j$ and $h(T_i, T_j) = T_i - T_j$. Then 
 $\widehat{\kappa}( L_i-L_j, T_i-T_j)$ can be viewed as an extension of the conditional production-moment correlation test  accommodating right censoring.
\end{example}

\begin{example}\label{ex2}
Consider a rank difference function,
%
$r(X_i, X_j) = {\rm rank}(X_i)/n - {\rm rank} (X_j)/n$, where ${\rm rank} (X_i)$ is the rank of $X_i$ among a dataset $\{X_i\}_{i=1}^{n}$. The function ${\rm rank}(\cdot)/n$ approximates the cumulative distribution function of $X$, $F_X$.
Then $\widehat{\rho}\{r(L_i, L_j),r(T_i, T_j)\}$ and $\widehat{\kappa}\{r(L_i, L_j),r(T_i, T_j)\}$ correspond to Spearman's rank correlation coefficient adapted to left-truncated data and to left-truncated and right-censored data, respectively, measuring the correlation among comparable pairs. The proposed test is likely to inherit advantages of  Spearman's rank correlation compared for non-truncated data such as effectively  detecting nonlinear dependence between two variables. 
\end{example}

\begin{example}\label{ex3}
One can also consider a hybrid of the Kendall's tau and the rank correlation, $\widehat{\rho}\{r(L_i, L_j), \sign(X_i-X_j)\}$ and $\widehat{\kappa}\{r(L_i, L_j), \sign(T_i-T_j)\}$. An advantage of the hybrid statistic $\widehat{\kappa}\{r(L_i, L_j), \sign(T_i-T_j)\}$  does not require Assumption 3B, as  the rank correlation $\widehat{\kappa}(r(L_i, L_j),r(T_i, T_j)\}$ in Example \ref{ex2} does.
\end{example}

Table \ref{table:unifycomp} summarizes existing approaches and the new tests according to the choice of $g$ and $h$, where ``sign'', ``rank'' and ``linear'' functions in the second and third column  are defined by $\sign(s-t)$, ${\rm rank}(s)/n - {\rm rank}(t)/n$, and $s-t$, respectively, for real $s$ and $t$. The three dependence measures on bivariate data,  Kendall's tau, Spearman's rank correlation, and Pearson's product-moment correlation, are systemically adjusted in our proposed test statistics.

\begin{table}[h]
\centering{\small
{
\begin{tabular}{cccl}
\hline
Data (statistic) & $g(\cdot, \cdot)$  & $h(\cdot, \cdot)$ & Equivalent test statistic \\ 
\hline
Left-truncated
	& Sign & Sign & \cite{Tsai1990}, $\widehat{\rho}^{\rm CTau}$ \\
($\widehat{\rho}(g,h)$)
	& Sign & Linear & A subclass of \cite{Jones1992}, \\
	&		&		& $\quad$ $T^{\rm Cox}(q,Z)$ with $q(t) = Y(t)$, $Z_i(t) = L_i$ \\
	& Linear & Linear & \cite{Chen1996}, $\widehat{\rho}^{\rm CProd}$ \\
	& Rank & Sign & (Example 3) \\	
	& Rank & Rank & (Example 2) \\	
\hline
Left-truncated
	& Sign & Sign & \cite{Tsai1990}, $\widehat{\kappa}^{\rm CTau}$ \\
and right-censored
	& Linear & Sign & A subclass of \cite{Jones1992},  \\
($\widehat{\kappa}(g,h)$)
	&		&		& $\quad$ $T^{\rm Cox}(q,Z)$ with $q(t) = Y(t)$, $Z_i(t) = L_i$ \\
	& Linear & Linear$^*$ & (Example 1) \\
	& Rank & Sign & (Example 3)\\	
	& Rank & Rank$^*$ & (Example 2) \\		
\hline
\end{tabular}
\caption{Special cases of the proposed class. The ``sign'', ``linear'', and ``rank'' functions denote bivariate mappings $\sign(s-t)$, $s-t$, and $r(s, t)$ defined in Section \ref{subsec:examples}, respectively. Tests requiring additional Assumption 3B are marked with asterisk($^*$).}
\label{table:unifycomp}
}
}
\end{table}

\section{Asymptotic relative efficiencies}\label{sec:AREs}

The fact that we can specify $g$ and $h$ leads to a natural question, which choice of $g$ and $h$ would be suitable to improve power to reject the quasi-independent hypothesis.
For complete bivariate data,
\cite{Randles1991} reported that Spearman's rank correlation and Kendall's tau had asymptotically comparable powers. 
The asymptotic relative efficiency of the rank correlation to the product-moment correlation was 0.91 when the assumptions for the latter were satisfied \citep{Daniel1990,Siegel1988}.
For left-truncated survival data, \cite{Jones1992} reported that $T^{\rm Cox}(q,L)$ with suitable choice of weight function $q$ is more efficient than the conditional Kendall's tau test. They used \cite{Jones1990}'s formula for asymptotic power under local contiguous alternatives.

We use the approach of  \cite{Jones1990} and \cite{Jones1992} to compare the theoretical powers of our proposed test statistics. If $h$ is the sign function, the proposed $\widehat{\rho}(g,\sign)$ and $\widehat{\kappa}(g,\sign)$ has equivalent form to \cite{Jones1992}'s test statistic, $T^{\rm Cox}$, as seen in \eqref{eqn:coxU1} and \eqref{eqn:coxU2} for several choices of $g$. We restrict our interest to the case of $g(L_i,L_j)$ being one of the followings: (1) $\sign(L_i-L_j)$ (``sign''); (2) $r(L_i, L_j)$ which is eventually approximated to $F_L(L_i) - F_L(L_j)$ (``rank''); and (3) $L_i-L_j$ (``linear''). We consider a sequence of contiguous hazard alternatives,
\[
H_{1}^n: \lambda(t \,|\, L_i ) = \lambda_1 (t) 
	\left\{ \alpha_1(t) + n^{-1/2} a(L_i) \beta + O(n^{-1}) \right\},
\]
where $a(\cdot)$ is a real-valued measurable function. The given alternative has been called as a relative risk model if $\alpha_1(t) = 1$ and an excess risk model if $\lambda_1(t) = 1$.
Following similar arguments to \cite{Jones1990},  we can show that the three test statistics  converge in distribution to the normal distribution with mean $\mu(\infty)/\sigma(\infty)$ and variance 1, where $\mu(t)$ and $\sigma(t)^2$ are calculated as in Table \ref{table:ARE1}. We introduce further notations in the table: $\bar{y}(t)$ is the limit of proportion of subjects at risk at time $t$; $\sigma_{XY}(t)$ is the limit of covariance of $X_i(t)$'s and $Y_i(t)$'s at risk at time $t$ for two processes $X(t)$ and $Y(t)$.
For completeness, we present the definitions stated in \cite{Jones1992}:
\begin{gather}
\sigma_{XY}(t) = \int_0^t X(l)Y(l) f_t(l)dl - \left(\int_0^t X(l) f_t(l)dl \right) \left( \int_0^t Y(l) f_t(l)dl \right) ,\NN \\
\mbox{where~} f_t(l) = I(l<t) 
	\frac{ \exp\left\{ - \int_0^l \Gamma_0(x)dx - \int_l^t \Gamma_1(x)dx \right\} \theta(l) \exp\left\{ - \int_0^l \theta(x)dx \right\} } { \int_0^t \exp\left\{ - \int_0^l \Gamma_0(x)dx - \int_l^t \Gamma_1(x)dx \right\} \theta(l) \exp\left\{ - \int_0^l \theta(x)dx \right\} dl }
	\label{eqn:waittime}
\end{gather}
with $\Gamma_i(x) = \psi_i(x) + \lambda_i(x)$ for $i=0,1$. Here, $\theta(t)$ is the hazard function of $L$, $\lambda_0(t)$ is the hazard rate of failure time of subjects who did not entered the study, and $\psi_1(t)$ and $\psi_0(t)$ are the hazard rates of censoring times of subjects who did and did not enter the study. Another note is that $\bar{y}(t)$ is the determinator of \eqref{eqn:waittime}. 
The efficacy of a given test is defined as $\mu(\infty)^2 / \sigma(\infty)^2$. We will compare the efficacies of the tests by Pitman asymptotic relative efficiency, the ratio of the efficacy of a test to another.

\begin{table}[h]
\centering
\begin{tabular}{cccc}
\hline
\multicolumn{2}{c}{$\widehat{\kappa}(g,h)$} & Equivalent form &  $\mu(t)$ and $\sigma^2(t)$ \\ \cline{1-2}
$g(\cdot, \cdot)$ & $h(\cdot, \cdot)$ & in $T^{\rm Cox}(q,Z)$ &  \\ \hline
Sign & Sign & $T^{\rm Cox}(Y,R^*)$ & $\mu(t) = \beta \int_0^{t}\bar{y}^2(u) \sigma_{a(L)R^*}(u) \lambda_1(u) du$  \\
& & & $\sigma(t)^2 = \int_0^{t}\bar{y}(u)^3 \sigma_{R^*R^*}(u) \lambda_1(u) \alpha_1(u) du$ \\ \hdashline
Rank & Sign & $T^{\rm Cox}(Y,F_L(L))$ & $\mu(t) = \beta \int_0^{t}\bar{y}(u)^2 \sigma_{a(L)F(L)}(u) \lambda_1(u) du$  \\ 
& & & $\sigma(t)^2 = \int_0^{t}\bar{y}(u)^3 \sigma_{F(L)F(L)}(u) \lambda_1(u) \alpha_1(u) du$ \\ \hdashline
Linear & Sign & $T^{\rm Cox}(Y,L)$ & $\mu(t) = \beta \int_0^{t}\bar{y}(u)^2 \sigma_{a(L)L}(u) \lambda_1(u) du$  \\
& & & $\sigma(t)^2 = \int_0^{t}\bar{y}(u)^3 \sigma_{LL}(u) \lambda_1(u) \alpha_1(u) du$ \\ \hline
\end{tabular}
\caption{Asymptotic efficacies for the select cases of the proposed class. Notations are defined in the main body of the paper.}
\label{table:ARE1}
\end{table}

We considered the following submodels of the contiguous alternatives:
\BIT
\item Model 1 (M1): $\lambda(t \,|\, L_i ) = 0.3 \cdot(1 + n^{-1/2} L_i \beta)$,
\item Model 2 (M2): $\lambda(t \,|\, L_i ) = 0.3 \cdot \left\{1 + n^{-1/2} (L_i^2 + \sin L_i)^{-1} \beta \right\}$,
\EIT
Both models are special cases of a relative risk or an excess risk model; M1 was selected to consider a model linear in $L$ and M2 was for one nonlinear in $L$. We generated $L$ from either the exponential distribution with rate 2 or the uniform distribution on $[0,1]$.
We chose $\lambda_0(t) = 0.3$ and $(\psi_0(t), \psi_1(t))$ as $(0, 0)$, $(0, 1)$, and $(1, 1)$ following \cite{Jones1992}. If $\psi_1(t) = 0$, there is no censoring on the observed patients and the resulting efficacies corresponds to those from the proposed class with no censoring. Similarly, by  $\psi_1(t) = 1$ we can compare the efficacies from censored data. In those settings, we could calculate the given $\mu(t)$'s and $\sigma(t)^2$'s under M1 by simple algebra and formulas given in Appendix 2 in \cite{Jones1992}. For M2, we approximated them by numerical integration.

Table \ref{table:ARE2} displays asymptotic relative efficiencies to $\widehat{\kappa}\{\sign(L_i-L_j), \sign(T_i-T_j)\}$ that is equivalent to the conditional Kendall's tau. We first focus on the case of $L \sim {\rm Exp}(2)$. The choice of $g$ as the rank function, $g(L_i, L_j) = F_L(L_i) - F_L(L_j)$,  performed better than and comparable to the conditional Kendall's tau when under M1 and M2, respectively. For the cases of $g(L_i, L_j) = L_i - L_j$ (the linear function), it performed the best under M1 where the dependence is linear and worst under M2 that depends on $L$ nonlinearly. When $L$ was uniform on $[0,1]$, we have $F_L(L) = L$ and the efficacies from $g$ as the rank and linear functions are theoretically the same. Their performance was slightly better than or similar to that of the conditional Kendall's tau.
From this limited comparison, we could see that the choice of $g$ as the linear function may lead to more powerful test under linear relationship between $L$ and $T$, but lead to a poor performance under certain nonlinear alternatives. On the other hand, the choice of $g$ as the rank function  performed better in some nonlinear cases.

\begin{table}[h]
\centering
\begin{tabular}{cccccccccc}
\hline
Model & $g(\cdot,\cdot)$ & $h(\cdot,\cdot)$ & \multicolumn{7}{c}{True distribution} \\
\cline{4-10}
 & &  & $L$ & \multicolumn{3}{c}{${\rm Exp}(2)$} & \multicolumn{3}{c}{${\rm Unif}(0,1)$} \\
 & & & $(\psi_0, \psi_1)$ & $(0,1)$ & $(0,1)$ & $(1,1)$ & $(0,0)$ & $(0,1)$ & $(1,1)$ \\ \hline
M1 & Rank & Sign & &  1.162 & 1.210 & 1.325 & 0.998 & 1.047 & 1.116 \\
& Linear & Sign & & 1.721 & 1.800 & 1.769 & 0.998 & 1.047 & 1.116  \\ \hline
M2 & Rank & Sign & & 1.028 & 1.039 & 1.010 & 1.001 & 1.008 &  1.018  \\
& Linear & Sign & & 0.402 & 0.401 & 0.414 & 1.001 & 1.008 &  1.018  \\ \hline
\end{tabular}
\caption{Asymptotic relative efficiencies of selected $\widehat{\kappa}(g,h)$'s relative to  $\widehat{\kappa}\{\sign(L_i-L_j),\sign(T_i-T_j)\}$ (Kendall's tau).}
\label{table:ARE2}
\end{table}

\section{Simulation study}\label{sec:simul}

We evaluated finite sample performances of the proposed test statistics $\widehat{\rho}(g,h)$ and $\widehat{\kappa}(g,h)$ with $g$ and $h$ chosen as in Table \ref{table:unifycomp}. 
%
Simulation scenarios mimic those of \cite{Jones1992} and \cite{Chen1996}. 

Pairs of $(L,X)$ were generated from the following two null and three non-null scenarios. 
We considered three exponential models where $L$ is uniformly generated from $[0,5]$ with the hazard function $x$ as $h(x|L) = 0.3$, $h(x|L) = 0.3 \, (1 - L/12)$, and $h(x|L) = 0.3 \, \{(L-2.5)^2 + 2\}^{-1}$, respectively. The first scenario represents the null case, the following two represent linear and nonlinear alternative cases. We also considered two normal models where $(L, X)^T$ follows a multivariate normal distribution with mean $(-1, 0)^T$ and covariance matrices as $[1, \rho ; \rho, 1]$, for $\rho=0, 0.15$, representing null and alternative cases, respectively.

For left-truncated and right-censored data, censoring time $C$ was independently generated from an exponential distribution, keeping the censoring rate around 40\%.  Finally, observations with $L \geq \min(X,C)$ were discarded. Sample size after the truncation was set to 400. For the generated dataset, we calculated the five $\widehat{\rho}(g,h)$'s and $\widehat{\kappa}(g,h)$'s with the choices of $g$ and $h$ presented in Table \ref{table:unifycomp} and conducted the quasi-independence hypothesis testing at the significance level 5\% according to the asymptotic null distribution shown in the Theorems.

Table \ref{table:simulation} reports empirical rejection rates over 5000 replications for the tests under the  aforementioned five null and alternative scenarios.   Under the null,  all the tests showed the rejection rates close to the nominal level. Under the alternative, there is no uniformly most powerful test, but depending on the alternatives, different tests had higher  power than the others.  
First, the conditional product-moment correlation was the most powerful under the normal alternative and the exponential alternative where the relationship between $L$ and $X$ was linear.   The newly proposed conditional Spearman's rank correlation featured in Example \ref{ex2} was the most powerful under the exponential alternative where the relationship between $L$ and $X$ was nonlinear. The results were consistent for both left-truncated and left-truncated and right-censored data. 
When we restricted comparisons among the tests not requiring Assumption 3B for left-truncated and right-censored data, the hybrid test with $g$ as linear and $h$ as sign, was the most powerful under the normal alternative and the exponential alternative with the linear link.  The new test with $g$ as rank and $h$ as sign featured in Example \ref{ex3} was the most powerful under the exponential alternative with nonlinear link and as powerful as the conditional Kendall's tau. These findings are consistent with those in Section \ref{sec:AREs}. Summarizing the results, newly proposed tests can be more powerful  than existing ones  under certain sets of alternatives.  



\begin{table}[h]
\centering{\small
{
\begin{tabular}{cccccccc}
\hline
Data (statistic) & $g(\cdot,\cdot)$ & $h(\cdot,\cdot)$ & \multicolumn{5}{c}{Empirical rejection frequency}  \\
\cline{4-8}
& & & \multicolumn{2}{c}{Null} & \multicolumn{3}{c}{Alternative} \\
 &   &   &   Exp.  & Norm. & Exp. (L) & Exp. (NL)  & Norm. \\ 
\hline
Left-truncated 
 & Sign & Sign   & 0.047 & 0.055 & 0.573 & 0.168 & 0.375 \\ 
($\widehat{\rho}(g,h)$)
 & Linear & Sign  & 0.043 & 0.052 & 0.641 & 0.104 & 0.412 \\
 & Linear & Linear  & 0.046 & 0.051 & 0.736 & 0.087 & 0.442 \\ 
 & Rank & Sign  & 0.043 & 0.053 & 0.628 & 0.177 & 0.383 \\ 
 & Rank & Rank  & 0.044 & 0.055 & 0.587 & 0.280 & 0.375 \\ 
\hline
Left-truncated
 & Sign & Sign  & 0.049 & 0.049 & 0.250 & 0.291 & 0.279 \\ 
and right-censored
  & Linear & Sign  & 0.050 & 0.050 & 0.343 & 0.162 & 0.300 \\   
($\widehat{\kappa}(g,h)$)
 & Linear & Linear*  & 0.047 & 0.051 & 0.384 & 0.115 & 0.314 \\
 & Rank & Sign  & 0.049 & 0.051 & 0.298 & 0.315 & 0.278 \\ 
 & Rank & Rank*  & 0.050 & 0.050 & 0.240 & 0.453 & 0.266 \\ 
\multicolumn{3}{c}{(freq. of censored data)}   & 0.401 & 0.401 & 0.402 & 0.401 & 0.400 \\
\hline
\end{tabular}
\caption{Empirical rejection rates of the test statistics $\widehat{\rho}(g,h)$ and $\widehat{\kappa}(g,h)$ with 5,000 simulated datasets from two null and three alternative scenarios, with $g(\cdot,\cdot)$ and $h(\cdot,\cdot)$ chosen according to Table \ref{table:unifycomp}. Tests requiring Assumption 3B are marked with asterisk ($^*$). Abbreviations: Exp., exponential model; Norm., normal model; L, linear; NL, nonlinear.}
\label{table:simulation}
}
}
\end{table}

\section{Data example: Channing House}\label{sec:data}

We applied the proposed quasi-independence tests to the Channing House data \citep{Hyde1980} that include ages at death of 97 males and 365 females who were residents of the Channing House retirement center  from January 1964 to July 1975. The data are left-truncated since a subject who died before  entering the community could not be recruited. In addition, the ages at death were right-censored by the end of the study.

Table \ref{table:realdata} shows the $p$-values of the five $\widehat{\kappa}(g,h)$'s. The tests with $h$ as the sign function show that the quasi-independence is marginally significant among the male group but not significant among the female group. The other two  tests where $h$ is not the sign function show a strong association between ages at  entrance and death.  Since these tests are valid under Assumption 3B, that is quasi-independence between censoring and truncation times, we inspected validity of this assumption by 
 reversing the role of survival time and censoring times and applying the tests $\widehat{\kappa}(g,h)$ with $h$ as the sign function. Table \ref{table:realdata2} displays corresponding $p$-values and suggests that the quasi-independence between $L$ and $C$ can be rejected.   Based on this finding, we adopted the results from the tests  which does not require Assumption 3B, and concluded that among women, we fail to reject that  failure time and truncation time are  quasi-independent.

\begin{table}[h]
\centering{\small
{
\begin{tabular}{cccc}
\hline
\multicolumn{2}{c}{$\widehat{\kappa}(g,h)$} & \multicolumn{2}{c}{Evaluated test statistic ($p$-value)}  \\ 
\hline
$g(\cdot, \cdot)$ & $h(\cdot, \cdot)$  & Men & Women \\ %
\hline
Sign & Sign & $3.972$ ($0.046$) & $0.600$ ($0.438$) \\
Linear & Sign & $3.248$ ($0.072$) & $0.663$ ($0.416$) \\
Linear & Linear & $7.142$ ($0.008$) & $11.682$ ($0.001$) \\
Rank & Sign & $3.749$ ($0.053$) & $0.521$ ($0.469$) \\
Rank & Rank & $7.315$ ($0.007$) & $8.287$ ($0.004$) \\
\hline
\end{tabular}
\caption{$p$-values of the test statistic $\widehat{\kappa}(g,h)$ with the Channing house datasets for testing truncation time and failure time.}
\label{table:realdata}
}
}
\end{table}

\begin{table}[h]
\centering{\small
{
\begin{tabular}{cccc}
\hline
\multicolumn{2}{c}{$\widehat{\kappa}(g,h)$} & \multicolumn{2}{c}{Evaluated test statistic ($p$-value)}  \\ 
\hline
$g(\cdot, \cdot)$ & $h(\cdot, \cdot)$  & Men & Women \\ 
\hline
Sign & Sign & $5.380$ ($0.020$) & $30.213$ ($< 10^{-7}$) \\
Linear & Sign & $7.490$ ($0.006$) & $37.393$ ($< 10^{-7}$) \\
Rank & Sign & $7.199$ ($0.007$) & $35.514$ ($< 10^{-7}$) \\
\hline
\end{tabular}
\caption{$p$-values of the test statistic $\widehat{\kappa}(g,h)$ with the Channing house datasets for testing truncation time and censoring time, by reversing the role of survival time and censoring time.}
\label{table:realdata2}
}
}
\end{table}

\section{Concluding remarks}\label{sec:conclusion}

We proposed a general class of tests which can embed existing tests for  quasi-independence of truncation time and survival time. The proposed class was built upon common characteristics of existing tests, namely, $U$-statistics of skew-symmetric transforms of all the pairs of comparable observations.  
For left-truncated and right-censored data, a subclass of tests in which  $h(\cdot,\cdot)$ is not the sign function require an additional  assumption of  quasi-independence between truncation and censoring times.  This subclass can be used after testing quasi-independence between truncation and censoring times via the proposed test with $h(\cdot,\cdot)$ as the sign function  exchanging the role of failure and censoring times. We also compared the powers from several choices of $g$ and $h$ theoretically and empirically. Our results suggest that  the choice of the linear function  may give a good power when the data has linear relationship, and a new sets of tests utilizing the rank function can be powerful under certain nonlinear alternatives.  The results resonate with those on Pearson's product-limit correlation, Spearman's rank correlation, and Kendall's tau in complete bivariate data.   Further research is needed to better understand when which choices of $g$ and $h$ would be suitable in variety of settings not considered in this paper. 



\section*{Appendix}
\subsection*{A. Proofs of the main results}

\begin{proof}[{\bf Proof of \eqref{eqn:coxU1}}]
Observe that
\begin{eqnarray}
T^{\rm Cox}\{Y, a(L)\} &=&
	 \sum_{i=1}^{n} \delta_i \, Y(T_i) \left\{ a(L_i) - \frac{ \sum_{j=1}^{n} Y_j(T_i) a(L_j) }{ Y(T_i) } \right\} \NN \\
&=&  \sum_{i=1}^{n} \delta_i \, \sum_{j=1}^{n} Y_j(T_i) \{a(L_i) - a(L_j)\} \NN \\
&=&  \sum_{i=1}^{n} \sum_{j=1}^{n} \delta_i  \, I(L_j<T_i) I(T_i \leq T_j) \{a(L_i) - a(L_j)\}. \label{eqn:coxreform}
\end{eqnarray}
 For further simplification, we denote $Q(i,j) = \delta_i  \, I(L_j<T_i) I(T_i \leq T_j)$.  We see that $Q(i,i)=0$, and $I(\Lambda_{ij}) = 1$ is equivalent to either $Q(i,j) = 1$ or $Q(j,i)=1$ for $i\neq j$.   Also,  without ties, $I(T_i \leq T_j) = I(T_i < T_j) = - I(T_i < T_j) \sign(T_i - T_j)$.  
These identities leads  \eqref{eqn:coxreform} to 
\begin{eqnarray}
T^{\rm Cox}\{Y, a(L)\} &=&
	 - \sum_{i=1}^{n} \sum_{j=1}^{n} \{a(L_i) - a(L_j)\}\sign(T_i - T_j)Q(i,j) \NN \\
&=& - \sum_{i=1}^{n-1} \sum_{j=i+1}^{n} \{a(L_i) - a(L_j)\}\sign(T_i - T_j)I(\Lambda_{ij}). \NN
\end{eqnarray}
\end{proof}

\begin{proof}[{\bf Proof of \eqref{eqn:coxU2}}]
Recalling the definition of $R_i(t)$, one notes that $\sum_{i=1}^{n} Y_i(t) R_i(t) = 1 + \cdots + Y(t) = Y(t)\{Y(t)+1\}/2$. In addition, $2 I(L_i<L_j) - 1 = -\sign(L_i - L_j)$ from the no-tie assumption. Then we have
\begin{eqnarray}
T^{\rm Cox}(Y, R^*) &=&
	 \sum_{i=1}^{n} \delta_i \, Y(T_i) \left\{ \frac{R_i(T_i)}{Y(T_i)} - 
	 \frac{ \sum_{j=1}^{n} Y_j(T_i) R_j(T_i) }{  Y(T_i)^2 } \right\} \NN \\
&=&  \frac{1}{2}\sum_{i=1}^{n} \delta_i \, \left\{ 2 R_i(T_i) - Y(T_i) - 1 \right\} \NN \\
&=&  \frac{1}{2}\sum_{i=1}^{n} \delta_i \, \left[ \sum_{j=1}^{n} Y_j(T_i) \big\{ 2 I(L_i<L_j) - 1 \big\} + 1\right] \NN \\
&=&  - \frac{1}{2}\sum_{i=1}^{n} \sum_{j=1}^{n} \delta_i \, 
	I(L_j<T_i) I(T_i \leq T_j) \sign(L_i-L_j)  + \frac{1}{2}\sum_{i=1}^{n} \delta_i. \label{eqn:coxreform2}
\end{eqnarray}
The left term in \eqref{eqn:coxreform2} has similar form with \eqref{eqn:coxreform} and it is easy to see 
$\sum_{i=1}^{n} \sum_{j=1}^{n} \delta_i \, I(L_j<T_i) I(T_i \leq T_j) \sign(L_i-L_j) =  \sum_{i=1}^{n-1} \sum_{j=i+1}^{n} \sign(L_i - L_j)\sign(T_i - T_j)I(\Lambda_{ij})$.
\end{proof}

\begin{proof}[{\bf Proof of Theorem \ref{thm1}}]

Let Assumptions \ref{A1} and \ref{A2}, and the quasi-independence hypothesis hold throughout the proof. For simplicity, we consider continuous random variables so that $(L,X)$ has a joint density $f_{L,X}(l,x)$.

By the central limit theorem for one-sample $U$-statistics \citep{Randles1991}, as $n \rightarrow \infty$,
\[
\sqrt{n}(\widehat{\rho}(g, h) - \rho (g, h)) \stackrel{d}{\longrightarrow}
	\mathcal{N} \left( 0, \frac{4\zeta}{{\rm pr}(\Omega_{12})^2}\right),
\]
where $\zeta = \phi(g,h)- [\rho (g, h) {\rm pr}(\Omega_{12})]^2$. The theorem holds if $\rho( g,h) = 0$. We claim \begin{equation}\label{eqn:proof1}
E\big[ g(L_1, L_2) h(X_1, X_2) I(\Omega_{12}) \big] = 0,
\end{equation}
which means that the numerator of $\rho( g,h)$ vanishes. 
Partition the event $\Omega_{12}$ into four disjoint events:  $A_1= ( L_1 < L_2 < X_1 < X_2  )$, $A_2 = ( L_1 < L_2 < X_2 < X_1  )$, $A_3 = ( L_2 < L_1 < X_1 < X_2  )$, and $A_4 = ( L_2 < L_1 < X_2 < X_1  )$. Note that $I(\Omega_{12}) = I(A_1) + I(A_2) + I(A_3) + I(A_4)$. In addition, the quasi-independence implies that $f_{L,X}(l,x) = f_L(l)f_X(x) / \alpha$ on $\{ (l,x) : l < x \}$ for some $\alpha$, where $f_L$ and $f_X$ are marginal densities of $L$ and $X$, respectively. By the skew-symmetry of $g$, we have
\begin{eqnarray}
&& E\big\{ g(L_1 , L_2) h(X_1,  X_2) I(A_1) \big\} \NN \\
&=& \int_{l_1 < l_2 < x_1 < x_2}  g(l_1 , l_2) h(x_1,  x_2)
	f_{L,X}(l_1,x_1) f_{L,X}(l_2,x_2) d(l_1, l_2, x_1, x_2) \NN \\
&=& \frac{1}{\alpha^2}\int_{l_1 = 0}^{\infty} \int_{l_2 = l_1}^{\infty} \int_{x_1 = l_2}^{\infty}
  \int_{x_2 = x_1}^{\infty} 
  g(l_1, l_2) h(x_1, x_2) f_{X}(x_1)f_{L}(l_1) f_{X}(x_2)f_{L}(l_2)  dx_2 dx_1 dl_2 dl_1 \NN \\
&=& \frac{1}{\alpha^2}\int_{l_1 = 0}^{\infty} \int_{l_2 = l_1}^{\infty} \int_{x_1 = l_2}^{\infty}
  \int_{x_2 = x_1}^{\infty} 
  \{-g(l_2, l_1)\} h(x_1, x_2) f_{X}(x_1)f_{L}(l_1) f_{X}(x_2)f_{L}(l_2)  dx_2 dx_1 dl_2 dl_1 \NN \\
&\stackrel{l_1 \leftrightarrow l_2}{=}& \frac{1}{\alpha^2}\int_{l_2 = 0}^{\infty} \int_{l_1 = l_2}^{\infty} \int_{x_1 = l_1}^{\infty}
  \int_{x_2 = x_1}^{\infty} 
  \{-g(l_1, l_2)\} h(x_1, x_2) f_{X}(x_1)f_{L}(l_2) f_{X}(x_2)f_{L}(l_1)  dx_2 dx_1 dl_1 dl_2 \NN \\
&=& - \int_{l_2 < l_1 < x_1 < x_2}  g(l_1 , l_2) h(x_1,  x_2)
	f_{L,X}(l_1,x_1) f_{L,X}(l_2,x_2) d(l_1, l_2, x_1, x_2) \NN \\
&=& - E\big\{ g(L_1 , L_2) h(X_1,  X_2) I(A_3) \big\}, \NN
\end{eqnarray}
which implies $E\big[ g(L_1 , L_2) h(X_1,  X_2) \{I(A_1) + I(A_3)\} \big] = 0$. A similar argument holds for $A_2$ and $A_4$. Thus \eqref{eqn:proof1} holds and the theorem follows. \qed

\end{proof}

\begin{proof}[{\bf Proof of Theorem \ref{thm2}}]

The proof goes in a similar way with that for Theorem \ref{thm1}. The asymptotic normality is straightforward from the central limit theorem. Our goal is to show
\begin{equation}\label{eqn:proof2}
E\big[ g(L_1 , L_2) \sign(T_1,  T_2) I(\Lambda_{12}) \big] = 0
\end{equation}
which implies $\kappa(g,h)=0$. We continue to suppose Assumption \ref{A1} and \ref{A2} along with the quasi-independence. For convenience, let $(L,X,C)$ be continuous and has a joint density $f_{L,X,C}(l,x,c)$. 

First, suppose that Assumption 3A holds. We verify \eqref{eqn:proof2} by proving the two facts: (a) $E\big\{ g(L_1 , L_2) \sign(T_1 - T_2) I(\Lambda_{12}) \big\} = E\big\{ g(L_1 , L_2) \sign(X_1 - X_2) I(\Lambda_{12}) \big\}$; and (b) $E \{ g(L_1 , L_2)$ $h(X_1,  X_2) I(\Lambda_{12}) \} = 0$ for any skew-symmetric $g$ and $h$.

For (a), write $\Lambda_{12} = B_1 \cup B_2 \cup B_3 \cup B_4$, where the four disjoint events are defined as follows: $B_1 = (L_1 < L_2 < X_1 < T_2) \cap (\delta_1 = 1)$, $B_2 = ( L_1 < L_2 < X_2 < T_1) \cap (\delta_2 = 1)$, $B_3 = ( L_2 < L_1 < X_1 < T_2) \cap (\delta_1 = 1)$, and $B_4 = (L_2 < L_1 < X_2 < T_1) \cap (\delta_2 = 1)$.
Observe that on $B_1$, we have $X_1 = T_1 < T_2 < X_2$ which implies $\sign (T_2 - T_1) = \sign (X_2 - X_1) = 1$. Similar arguments hold for $B_2$, $B_3$, and $B_4$. Thus $E\big[ g(L_1 , L_2) \sign(T_1 - T_2) I(\Lambda_{12}) \big] = E\big[ g(L_1 , L_2) \sign(X_1 - X_2) I(\Lambda_{12}) \big]$.

For (b), combine Assumption \ref{A1}, \ref{A2}, and the quasi-independence hypothesis to obtain $f_{L, X, C}(l,x,c) = f_X(x)f_{L, C}(l,c) / \alpha$ on $\{(l,x,c) : l < x, l < c \}$ for some $\alpha$, where $f_X$ and $f_{L, C}$ are the marginal densities of $X$ and $(L, C)$, respectively. In addition, note that $B_1 = (L_1 < L_2 < X_1 < \min \{X_2, C_1, C_2\})$ and the other disjoint events can be written similarly. 
We now claim $E\big[ g(L_1 , L_2) h(X_1,  X_2) \{ I(B_1)  + I(B_3)\} \big] = 0$. Observe that
\begin{eqnarray}
&& E\big\{ g(L_1 , L_2) h(X_1,  X_2) I(B_1) \big\} \NN \\
	&=& \int_{ l_1 < l_2 < x_1 < \min \{x_2, c_1, c_2\}  } g(l_1, l_2) h(x_1, x_2) f_{L, X, C}(l_1,x_1,c_1) f_{L, X, C}(l_2,x_2,c_2) d(l_1, x_1, c_1, l_2, x_2, c_2) \NN \\
&=& \frac{1}{\alpha^2} \int_{l_1 = 0}^{\infty} \int_{l_2 = l_1}^{\infty} \int_{x_1 = l_2}^{\infty}
  \int_{x_2 = x_1}^{\infty} \int_{c_1 = x_1}^{\infty} \int_{c_2 = x_1}^{\infty}
  g(l_1, l_2) h(x_1, x_2) \cdot \NN \\
  && \qquad \qquad f_{X}(x_1)f_{L,C}(l_1,c_1) f_{X}(x_2)f_{L,C}(l_2,c_2) dc_2 dc_1 dx_2 dx_1 dl_2 dl_1 \NN \\
&=& \frac{1}{\alpha^2} \int_{l_1 = 0}^{\infty} \int_{l_2 = l_1}^{\infty} \int_{x_1 = l_2}^{\infty}
  g(l_1, l_2) f_{X}(x_1) {\rm pr}(x_1) Q(l_1,x_1) Q(l_2, x_1) dx_1 dl_2 dl_1 \NN
\end{eqnarray}
where $Q(l,x) = \int_{c = x}^{\infty} f_{L,C}(l,c) dc$ and ${\rm pr}(x) = \int_{u =x}^{\infty} h(x, u) f_{X}(u)du$.
On the other hand, the expectation on $B_3$ can be calculated by
\begin{eqnarray}
&& E\big\{ g(L_1 , L_2) h(X_1,  X_2) I(B_3) \big\} \NN \\
&=& \frac{1}{\alpha^2}  \int_{l_2 = 0}^{\infty} \int_{l_1 = l_2}^{\infty} \int_{x_1 = l_1}^{\infty}
  g(l_1, l_2) f_{X}(x_1) {\rm pr}(x_1) Q(l_1,x_1) Q(l_2, x_1) dx_1 dl_1 dl_2 \NN \\
&\stackrel{l_1 \leftrightarrow l_2}{=}& \frac{1}{\alpha^2}  \int_{l_1 = 0}^{\infty} \int_{l_2 = l_1}^{\infty} \int_{x_1 = l_2}^{\infty}
  g(l_2, l_1) f_{X}(x_1) {\rm pr}(x_1) Q(l_2,x_1) Q(l_1, x_1) dx_1 dl_2 dl_1 \NN \\
&=& \frac{1}{\alpha^2}  \int_{l_1 = 0}^{\infty} \int_{l_2 = l_1}^{\infty} \int_{x_1 = l_2}^{\infty}
  \left\{-g(l_1, l_2)\right\} f_{X}(x_1) {\rm pr}(x_1) Q(l_2,x_1) Q(l_1, x_1)
  dx_1 dl_2 dl_1 \NN \\
&=& - E\big\{ g(L_1 , L_2) h(X_1,  X_2) I(B_1) \big\},  \NN
\end{eqnarray}
where the skew-symmetry of $g$ was used at the third inequality.
Thus the claim holds and similarly $E\big[ g(L_1 , L_2) h(X_1,  X_2) \{ I(B_2)  + I(B_4)\} \big] = 0$, which implies \eqref{eqn:proof2}. \qed

Now we consider the case when Assumption 3B holds. Integrating Assumption \ref{A1}, \ref{A2}, and 3B, $f_{L, X, C}(l,x,c)$ is equal to $f_{L}(l)f_X(x)f_{C}(c) / \alpha$ on $\{(l,x,c) : l < x, l < c \}$ for some $\alpha$, where $f_L$, $f_X$, $f_C$ is the marginal densities of $L$, $X$ and $C$, respectively. Then
\begin{eqnarray}
&& E\big\{ g(L_1 , L_2) h(T_1,  T_2) I(B_1) \big\} \NN \\
%
&=& \frac{1}{\alpha^2} \int_{l_1 = 0}^{\infty} \int_{l_2 = l_1}^{\infty} \int_{x_1 = l_2}^{\infty}
  \int_{x_2 = x_1}^{\infty} \int_{c_1 = x_1}^{\infty} \int_{c_2 = x_1}^{\infty}
  g(l_1, l_2) h(x_1, \min(x_2, c_2) ) \cdot \NN \\
  && \qquad \qquad f_{X}(x_1)f_{L}(l_1)f_{C}(c_1) f_{X}(x_2)f_{L}(l_2)f_{C}(c_2) dc_2 dc_1 dx_2 dx_1 dl_2 dl_1 \NN \\
&=& \frac{1}{\alpha^2} \int_{l_1 = 0}^{\infty} \int_{l_2 = l_1}^{\infty} \int_{x_1 = l_2}^{\infty} \int_{x_2 = x_1}^{\infty}
  g(l_1, l_2) f_{X}(x_1)f_{L}(l_1) f_{X}(x_2)f_{L}(l_2) \cdot \NN \\
  && \qquad \qquad
  \Big\{ \int_{c_1 = x_1}^{\infty} f_{C}(c_1) dc_1 \Big\}
  \Big\{ \int_{c_2 = x_1}^{\infty} h(x_1, \min(x_2, c_2) )f_{C}(c_2) dc_1 \Big\}
  dx_2 dx_1 dl_2 dl_1 \NN
\end{eqnarray}
On the other hand, on $B_3$, interchanging $l_1 \leftrightarrow l_2$ and the skew-symmetry of $g$ yields
\begin{eqnarray}
&& E\big\{ g(L_1 , L_2) h(T_1,  T_2) I(B_3) \big\} \NN \\
&=& \frac{1}{\alpha^2} \int_{l_2 = 0}^{\infty} \int_{l_1 = l_2}^{\infty} \int_{x_1 = l_1}^{\infty} \int_{x_2 = x_1}^{\infty}
  g(l_1, l_2) f_{X}(x_1)f_{L}(l_1) f_{X}(x_2)f_{L}(l_2) \cdot \NN \\
  && \qquad \qquad
  \Big\{ \int_{c_1 = x_1}^{\infty} f_{C}(c_1) dc_1 \Big\}
  \Big\{ \int_{c_2 = x_1}^{\infty} h(x_1, \min(x_2, c_2) )f_{C}(c_2) dc_1 \Big\}
  dx_2 dx_1 dl_1 dl_2 \NN \\
&=& \frac{1}{\alpha^2} \int_{l_1 = 0}^{\infty} \int_{l_2 = l_1}^{\infty} \int_{x_1 = l_2}^{\infty} \int_{x_2 = x_1}^{\infty}
  g(l_2, l_1) f_{X}(x_1)f_{L}(l_2) f_{X}(x_2)f_{L}(l_1) \cdot \NN \\
  && \qquad \qquad
  \Big\{ \int_{c_1 = x_1}^{\infty} f_{C}(c_1) dc_1 \Big\}
  \Big\{ \int_{c_2 = x_1}^{\infty} h(x_1, \min(x_2, c_2) )f_{C}(c_2) dc_1 \Big\}
  dx_2 dx_1 dl_1 dl_2 \NN \\
&=& - E\big\{ g(L_1 , L_2) h(T_1,  T_2) I(B_1) \big\}.  \NN
\end{eqnarray}
The rest of the proof goes in a same way with the previous proofs and we have \eqref{eqn:proof2}. \qed

\end{proof}


\end{document}